\documentclass[12pt, draftcls, onecolumn]{IEEEtran}
\usepackage[cmex10]{amsmath}
\usepackage{subfigure}
\usepackage{amssymb}
\usepackage{amsthm}
\usepackage{epsfig}
\usepackage{color}
\usepackage{algorithm}
\usepackage{algorithmic}
\usepackage{multirow}
\usepackage{array}

\usepackage{epsfig}
\usepackage{color}

\begin{document}
\title{Analytical Models for Power Networks:\\ The case of the Western US and ERCOT grids}
\author{Deepjyoti Deka and Sriram Vishwanath\thanks{Deepjyoti Deka and Sriram Vishwanath are at The Lab. of Informatics, Networks and Communications, Department of Electrical and Computer Engineering, University of Texas at Austin. This work is supported by the Defense Threat Reduction Agency (DTRA) through grant \# HDTRA1-09-1-0048-P00003 }}

\maketitle

\begin{abstract}
The topological structure of the power grid plays a key role in the reliable delivery of electricity and price settlement in the electricity market. Incorporation of new energy sources and loads into the grid over time has led to its structural and geographical expansion and can affect its stable operation. This paper presents an intuitive analytical model for the temporal evolution of large grids and uses it to understand common structural features observed in grids across America. In particular, key graph parameters like degree distribution, graph diameter, betweenness centralities, eigen-spread and clustering coefficients, as well as graph processes like infection propagation are used to quantify the model's benefits through comparison with the Western US and ERCOT power grids. The most significant contribution of the developed model is its analytical tractability, that provides a closed form expression for the nodal degree distribution observed in large grids. The discussed model can be used to generate realistic test cases to analyze topological effects on grid functioning and new grid technologies.
%{Insert keyword text here.}
%%%% If classification number provided then
%\\
%2000 Math Subject Classification: 34K30, 35K57, 35Q80,  92D25
\end{abstract}

\section{Introduction}

The power grid refers to the complex network of energy generators, consumers and connecting transmission lines that are involved in the delivery of electricity. The advent of the `smart grid' has added further complexity to the system due to the introduction of new renewable generation resources and flexible consumer demand response. These advances have the potential to make the grid green as well as to reduce electricity pricing. To understand their impact correctly, such technologies need to be implemented on test cases that reflect the structure and functionality of current as well as future grids. Traditional power grid test cases used in literature are generally much smaller in size compared to real grids \cite{uwash}. From a topological perspective, there is thus a need to develop test cases that mirror the common trends in large grids around the world, and more so to provide an explanation for the temporal evolution of such common observations.

Modeling the power grid is a growing field of research. Past research has demonstrated that the performance of the power grid is dependent on its topology, normally represented by the adjacency matrix or the degree distribution. Albert et al. \cite{albert2000} have established the importance of the physical network of the power grid in determining its vulnerability to random and directed attacks. It is worth mentioning that there is a body of work categorizing the degree distribution observed in the underlying network of large power grids. \cite{albert2004exp2} reports the presence of exponential nodal degree distribution while considering power plants, substations, and 115 - 765 kV power lines of the North American power grid. The exponential degree distribution is also reported for the Western US power grid in \cite{amaral2000}. Hines et al. in \cite{hines2011} suggest an exponential degree distribution  in the Eastern, Western and Texas Interconnects in North America. A clear exponential tail distribution for the New York Independent System Operator (NYISO) is mentioned in \cite{wang2010}. Similarly, \cite{crucittia2004,sole2007a,sole2007b} mention an exponential degree distribution for power grids in several European countries, namely Italy, UK, Ireland, Portugal, and for the entire connected component in the Union for the Co-ordination of Transmission Electricity, UCTE. It is thus fair to state that despite differing geographical shapes and terrains, power grids across America and Europe possess an exponential distribution or, in the least, an exponential tail for the nodal degrees. This distinguishes power grids from other studied networks like social networks that tend to have power-law/scale free distributions.

Similarly, generative models for the observed exponential degree distribution in the power grid can be found in literature. \cite{watts1998} presents the well-known small-world generative model where nodes get connected to neighbors in a ring lattice and links get rewired between nodes to form the final graph. \cite{wang2010} develops a model called RT-nested-Smallworld based on the Watts and Strogatz small-world model. Nodes, in this model, lie on a ring lattice and are linked to their neighbors with some probability. These links are then rewired following a Markov chain to develop the final network. The model presented in  \cite{wang2010} approximates the NYISO power grid well, though it requires several tunable parameters to obtain the optimal fit. This results in the model yielding limited intuition regarding the origin of the observed network structure. One major drawback of such generative models, from an explanatory viewpoint, is that they presume a static original network configuration. In reality, power grids are dynamic and evolve with time due to addition of new nodes (buses) and edges (transmission lines). Further the aforementioned model does not describe the spatial nature of the power grid with physical nodes distributed over a geographical area. In related work, networks demonstrating a power-law nodal degree distribution like the internet and airline network have been explained using the Barabasi-Albert model \cite{barabasi}. \cite{pennock} presents a modified generative model based on the Barabasi-Albert model for exponentially distributed networks. Here, new nodes enter the network and gets connected to existing nodes either randomly or with a probability that gives preference to nodes with higher degrees. Besides the absence of spatial nodal placement, this model is restrictive as it uses the principle of `the rich gets richer' which clearly does not apply to power grids.

The primary goal of this paper is thus to provide a spatio-temporal generative framework that intuitively explains the common trends observed in evolving grids and in particular, provides an analytical expression for the observed nodal degree distribution. To demonstrate the efficacy of our approach, we also compare important features including diameter, node and edge betweenness, spread of eigenvalues and graph clustering of our generative model with real grids, notably the Western US and the ERCOT grids. Note that analytical models like ours, unlike detailed modeling and simulation tools, do not lead to an exact match for all characteristics observed in the real world. Instead, our model provides an accurate approximation for common structures with the additional benefit of mathematically tractability for important network measures. General information on the relative advantages and approximations associated with network modeling techniques can be found at \cite{sample}.

The rest of this paper is organized as follows. The next section presents the mathematical characterization of the empirical node degree distributions observed in two power grids in America. The generative framework for modeling the evolving grid is introduced in Section \ref{sec:gm} along with the theoretical analysis of the resulting node degree distribution. The correctness of the nodal degree distribution is shown through comparison with two power grids, the Western US and ERCOT grids, in Section \ref{sec:real fit}. The diameter calculations for our generative model for different sizes of the network are discussed in Section \ref{sec:diameter of the proposed model}. It is followed by a comparison of the node and edge centrality measures between real grids and generated networks in Section \ref{sec:Betweenness Centrality of the proposed model}. The distribution of node clustering coefficients  and spread of eigen-values for the available power grids are presented in Section \ref{sec:Node Clustering of the proposed model} and \ref{sec:eigen spread in the proposed model} respectively. Finally, dynamics of infection propagation processes on networks generated by our model and real grids are discussed in Section \ref{sec:Vulnerability of the proposed model}. Concluding remarks are summarized in Section \ref{sec:results and conclusion}.

\section{Characterizing the Empirical Node Degree Distribution}
\label{sec:emp}
\begin{figure}[!bt]
\centering
\subfigure[]{\includegraphics[width=.47\textwidth,height=.45\textwidth]{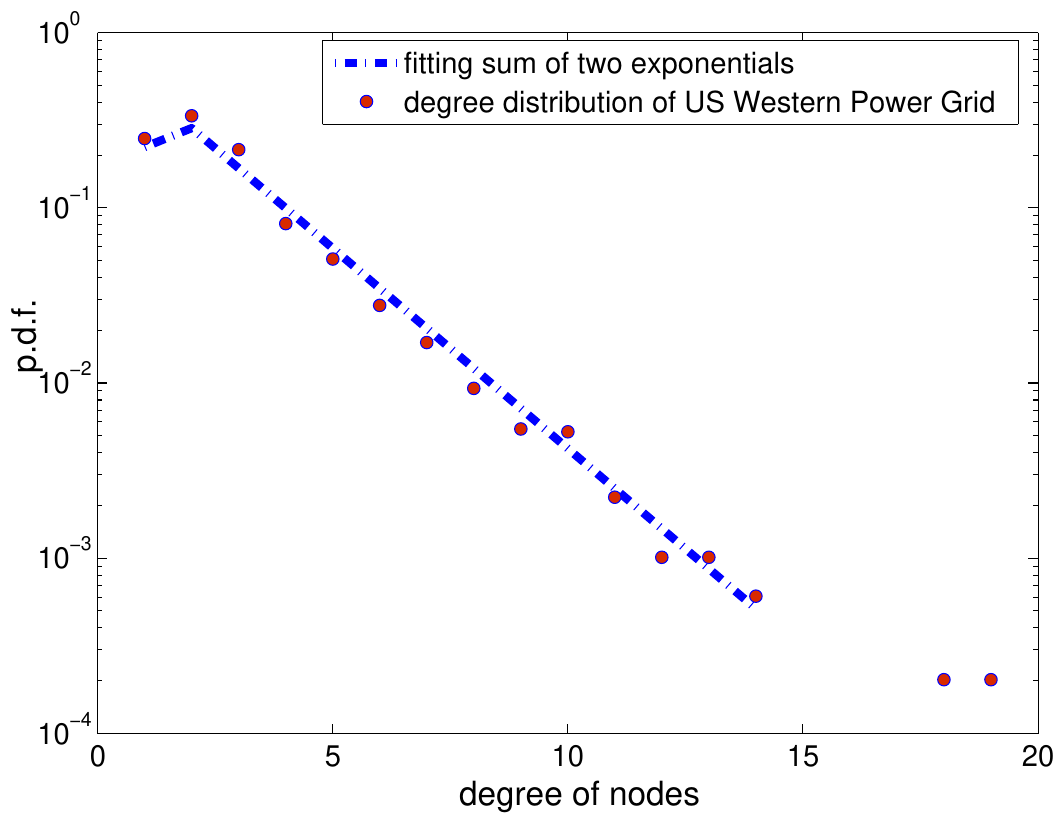}\label{fig:westernusagrid}}%\hspace{.05cm}
\subfigure[]{\includegraphics[width=.47\textwidth,height=.45\textwidth]{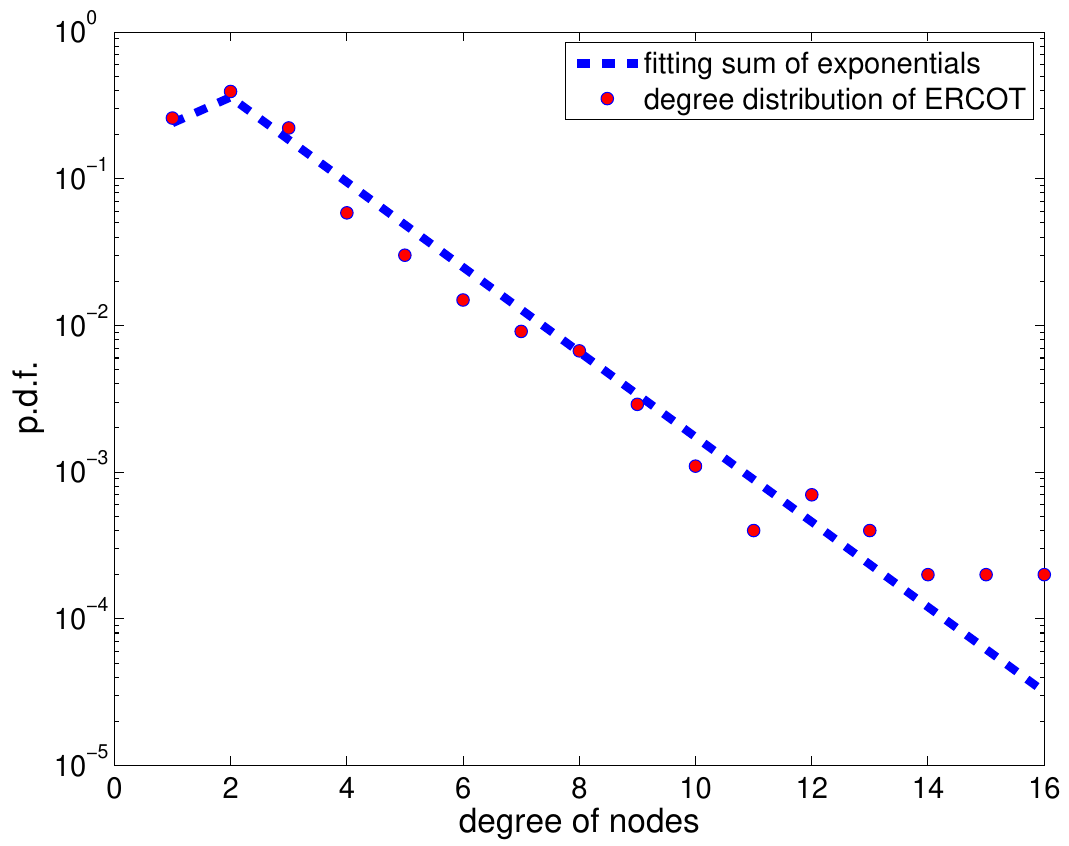}\label{fig:texasgrid}}%\hspace{.05cm}
\vspace{-.25cm}
\caption{Observed node degree distribution p.d.f in power grids (a) Western US (b) ERCOT}
%\label{fig:degreedist}
\vspace{-3mm}
\end{figure}

Structurally, the power grid is a graph with nodes representing the buses in the grid and edges representing the transmission lines connecting the buses. As mentioned in the Introduction, power grids across the world have been reported to have node degree distributions that have exponential tails. We consider three power grids in this paper, and evaluate the node degree distributions of their representative networks. Fig.~\ref{fig:westernusagrid} shows the node degree distribution of the Western US electrical power grid, which has $4941$ nodes and $6594$ edges. The network data for this grid is freely available \cite{watts1998}. The second network we consider is the power grid in Texas that is under the Electric Reliability Council of Texas (ERCOT) \cite{ercot}. The connected graph has $5514$ nodes and $6522$ edges and its node degree distribution is presented in Fig.~\ref{fig:texasgrid}.

We observe that for each power grid, the pd.f. of the degree distribution has a significant exponential tail. Additionally, we note that each considered degree distribution increases in value over a short interval preceding the exponential tail. The same trend has been reported for the NYISO power grid in \cite{wang2010}, where the increase at the beginning has been called a `kink'. Due to the presence of the initial interval of increase, the node degree distribution deviates from an exact exponential function. As elucidated in Figs.~\ref{fig:westernusagrid} and \ref{fig:texasgrid}, the empirical degree distribution is best fit by a \textbf{shifted sum of exponential distributions} characterized below:
\begin{align}
pdf(d) = \sum_i w_i e^{-\frac{d-k_i}{\mu_K}}\mathbb{I}(d \geq k_i) \label{dist}
\end{align}
where $d$ denotes the degree of a node and $w_i$ and $k_i$ are normalizing constants and degree shifts associated with each exponential. In the next section, we present our generative model that produces temporally evolving networks with degree distributions similar to that of real grids.

\section{Generative Model for Power Grid Network}
\label{sec:gm}
A brief note on notation: we use $P(.)$ to denote the probability of an event, and $E(.)$ to denote expectation. $\delta(.)$ represents the dirac delta function, ${\overline{\alpha}} = 1 - \alpha$ and ${\mathbb{I}}(.)$ the indicator function.

We show in the previous section that power grids possess an degree distribution that can be represented by a sum of shifted exponentials. Our goal here is to explain it using a temporally-evolving generative model based on the framework of 2-D spatial point process theory. In our model, $N_{total}$ nodes are placed one at a time in space according to a two dimensional Poisson point process $P_\lambda$\ with density $\lambda$. Here, edges/links between nodes are formed when a new node is placed into the system. The new node connects to $K$ of its nearest pre-existing nodes in the network, where $K$ is a parameter given by an integer-valued random variable with known distribution. The growth step captures the expansion of power grids over time due to the creation of new buses and resulting transmission lines. The parameter $K$ here represents the tendency of new nodes to form multiple connections to build robustness against failures in the physical network.  The selection of `nearest neighbors' for edge formation minimizes the costs of building transmission lines which require significant investment in real estate and copper. We now, prove that our model of network growth provides the appropriate shifted-exponential degree distribution observed in real power grids.

\subsection{Analysis of the Degree Distribution of the Generative Model}
\label{sec:mean field}
\begin{figure}[ht]
\centering
\includegraphics[width=.45\textwidth,height=.45\textwidth]{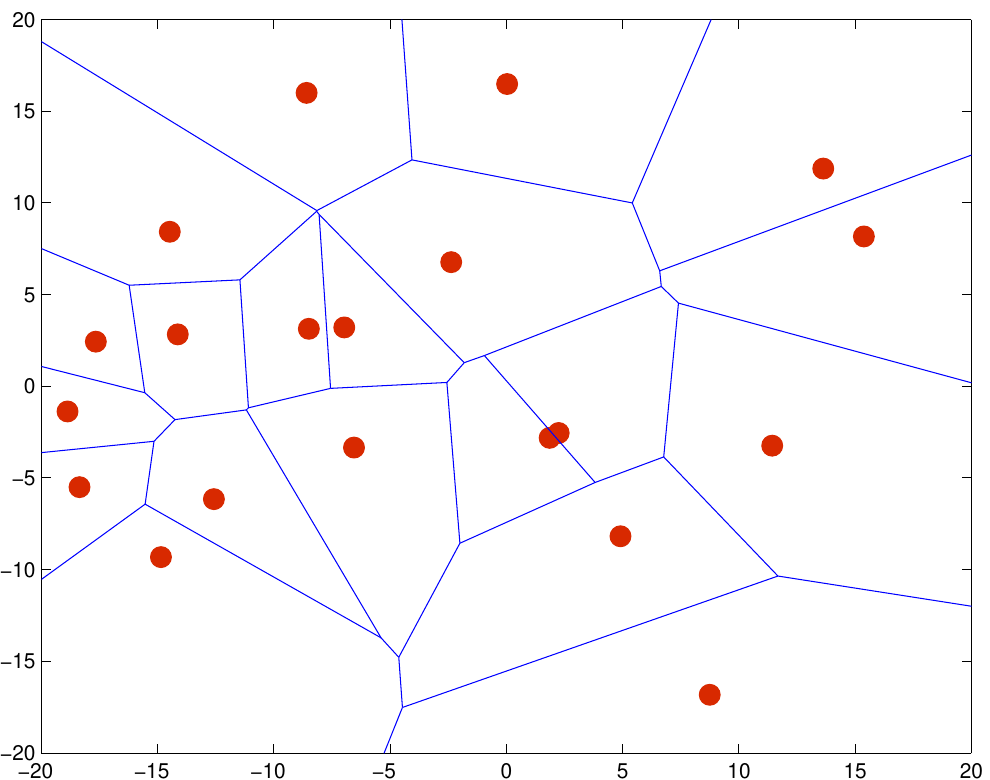}
\label{fig:voronoi}
\caption{Region of Influence represented by Voronoi region for a 20 nodes system with $K =1$}
\end{figure}

We represent the geographical area covered by the grid by a disk $R$. The network evolution in our generative model takes place in discrete time steps with one new node being born at each step. The nodes are numbered in the order of their creation with the first-born node numbered as $1$. From the network growth process described in Section \ref{sec:gm}, it is clear that a new node born at time $t$ connects to an existing node $a$ present in $R$ if its location (chosen randomly) lies within $a$'s `Region of Influence' ($R_{at}$) - the area within $R$ at time $t$ such that for every point $b$ within that area, node $a$ is amongst the $K$ nearest neighbors of $b$. As shown in Fig.~\ref{fig:voronoi}, for $K =1$ (new nodes link to their nearest neighbors), the `region of influences' of existing nodes are given by a first order Voronoi tessellation \cite{voronoi} in the disk $R$.

For $K > 1$, the regions of influence are given by overlapping $K^{th}$ order Voronoi-regions. The analysis of the degree distribution in the evolving network thus follows from the evolution of $K^{th}$ order Voronoi-regions defined by the existing nodes in $R$. As closed form expressions for higher order ($K >1$) Voronoi-regions seldom arise, we use a mean-field framework \cite{jackson} for our analysis instead. This technique involves replacing the dynamics of a single node at each step with the average dynamics of the entire system.

We first consider the case where $K$ (number of links formed by the new node) is a constant at $k$. Note that due to creation of $k$ edges, each new node gets a degree of $k$. Let $N(m,t)$ denote the average number of nodes of degree $m$ at time $t$. At time $t+1$, $N(m,t)$ increases if the new node born into the system connects with nodes of degree $m-1$, and decreases if the new node links with nodes of degree $m$. Now the probability of forming an edge to a node of degree $m$ is proportional to the cumulative volume under `regions of influence' of nodes of degree $m$. Note that at each time step, the expected volumes of the $k^{th}$ order Voronoi-regions of all existing nodes depend only on their positions (which are selected randomly) and not on the order of their arrival into the system. Subsequently, each existing node has the same expected volume of `region of influence'. Thereby the expected cumulative volume of the regions of influence of nodes with degree $m$ is proportional to the fraction of nodes with degree $m$. The average increase in the number of nodes of degree $m$ at a time step is thus given by the following relation

\begin{align}
N(m,t+1) - N(m,t) = & \frac{N(m-1,t)k}{t} - \frac{N(m,t)k}{t} + \delta(m=k)
\end{align}
Consider $m=k$. We have $N(k,t+1) - N(k,t) =  -\frac{N(k,t)k}{t} + 1$. For large $t$, using Lemma 4.1.1. in \cite{Durrett}, this becomes
\begin{align}
~\frac{N(k,t)}{t}  = \frac{1}{1+k} \label{equ1}
\end{align}
For $m \neq k$, we have
\begin{align}
N(m,t+1)  = N(m,t)(1-\frac{k}{t}) + \frac{N(m-1,t)k}{t}
\end{align}
For large $t$, again using Lemma 4.1.2 in \cite{Durrett}, we obtain
\begin{align}
&\frac{N(m,t)}{t}  = \frac{\lim_{t \rightarrow \infty}\frac{N(m-1,t)}{t}k}{1+k} \label{equ2}\\
\Rightarrow~&\frac{N(m,t)}{t}  = (\frac{k}{1+k})^{(m-k)}\frac{1}{1+k}
\end{align}
The average fraction of nodes $n(d)$ with degree less than $d$, where $d \geq k$, is given by
\begin{align}
&n(d) = \frac{1}{1+k} + \frac{1}{1+k}\frac{k}{1+k} + ... + \frac{1}{1+k}(\frac{k}{1+k})^{(d-k-1)} \nonumber \\
\Rightarrow~&n(d) = 1 - (\frac{1+k}{k})^{(k-d)}  \approx 1 - e^{-\frac{d- k}{k}}
\end{align}

Note that the cumulative distribution function converges to an exponentially decaying function for constant $K$. Next, we analyze the case where parameter $K$ is an integer valued bounded random variable with known distribution. Let $K$ take two values, $k_1$ and $k_2$ with probability $\alpha$ and $\overline{\alpha}$ respectively. Thus, $K$ has a mean of $\mu_{K} = \alpha k_1 + \overline{\alpha}k_2$ where $\overline{\alpha} = 1-\alpha$. Note that $\mu_{K}$ represents the average number of edges formed at each time step by a new node. The temporal evolution in expected number of nodes of degree $m$, $N(m,t+1)$ is then governed by the following equation:
\begin{align}
N(m,t+1) - N(m,t) =&~\mu_{K}\frac{N(m-1,t)}{t} - \mu_{K}\frac{N(m,t)}{t} + \alpha\delta(m=k_1) + \overline{\alpha}\delta(m=k_2)
\end{align}
As a fraction $\alpha$ of new nodes form $k_1$ links, we have
$$N(k_1,t+1) - N(k_1,t) =  -\frac{N(k_1,t)\mu_{K}}{t} + \alpha.$$

For large $t$, using (\ref{equ1}), this becomes
\begin{align}
~\frac{N(k_1,t)}{t}  = \frac{\alpha}{1+\mu_{K}}  \label{equ3}
\end{align}
For $k_1 < m < k_2$, at large $t$, using (\ref{equ2}), we have
\begin{align}
&\frac{N(m,t)}{t}  = \frac{\lim_{t \rightarrow \infty}\frac{N(m-1,t)}{t}\mu_{K}}{1+\mu_{K}} = (\frac{\mu_{K}}{1+\mu_{K}})^{(m-k_1)}\frac{\alpha}{1+\mu_{K}}   \label{equ4}
\end{align}
Similarly, for $m=k_2$, at large $t$,
\begin{align}
&\frac{N(k_2,t)}{t} = (\frac{\mu_{K}}{1+\mu_{K}})^{(k_2-k_1)}\frac{\alpha}{1+\mu_{K}} + \frac{\overline{\alpha}}{1+\mu_{K}} \label{equ5}
\end{align}
Finally, for $m > k_2$, for large $t$, using the same method we get:
\begin{align}
\frac{N(m,t)}{t} &= (\frac{\mu_{K}}{1+\mu_{K}})^{(m-k_1)}\frac{\alpha}{1+\mu_{K}} + (\frac{\mu_{K}}{1+\mu_{K}})^{(m-k_2)}\frac{\overline{\alpha}}{1+\mu_{K}} \label{equ6}
\end{align}
\\
Using Equations (\ref{equ3}), (\ref{equ4}), (\ref{equ5}) and (\ref{equ6}), the fraction of nodes $n(d)$ with degree less than $d$ is given by
\begin{align}
n(d) \approx \alpha(1 - e^{-\frac{d- k_1}{\mu_{K}}})\mathbb{I}(d > k_1) + \overline{\alpha}(1 - e^{-\frac{d- k_2}{\mu_{K}}})\mathbb{I}(d > k_2)
\end{align}

Using an identical approach, it is clear that if $K$ takes values $k_i$ with probability $\alpha_i$ respectively over some range of $i$'s, the nodal degree distribution is given by a weighted shifted sum of exponentials:
\begin{align}
pdf(d) = \sum_i\frac{\alpha_i}{\mu_K}e^{-\frac{d-k_i}{\mu_K}}\mathbb{I}(d \geq k_i) \label{varialbek}
\end{align}

\textbf{Validity of Mean-Field:} We now discuss the correctness of using a mean-field model to approximate the degree distribution in our analysis of the generative model. We first construct a random variable $Z(m,t)$  to represent the actual number of nodes of degree $m$ at time $t$. Let the locations of incoming nodes in the system be given by $y_1$, $y_2$ etc. beginning with the first node. We can then obtain a martingale $Z_i(m,t)$  as follows:
\begin{align}
Z_i(m,t)  = E(Z(m,t)|y_1,y_2...,y_{i})
\end{align}
Note that the average number of nodes of degree $m$ at time $t$, ~$N(m,t) = Z_0(m,t) = E(Z(m,t)$, while $Z(m,t) = Z_t(m,t)$. Further, $|Z_i(m,t)-Z_{i-1}(m,t)| \leq 2k_{max}$ where $k_{max}$ is the maximum value that $K$ can take. Using the Azuma-Hoeffding's inequality \cite{Durrett} for martingale convergence, we have
\begin{align}
& P(Z(m,t)-N(m,t)\geq x) \leq e^{-\frac{x^2}{8k^2t}} \nonumber\\
\Rightarrow~&P(Z(m,t)/t-N(m,t)/t\geq x) \rightarrow 0 \text{ as } t\rightarrow \infty\nonumber
\end{align}

\begin{figure}[!bt]
\centering
\subfigure[]{\includegraphics[width=.47\textwidth,height=.45\textwidth]{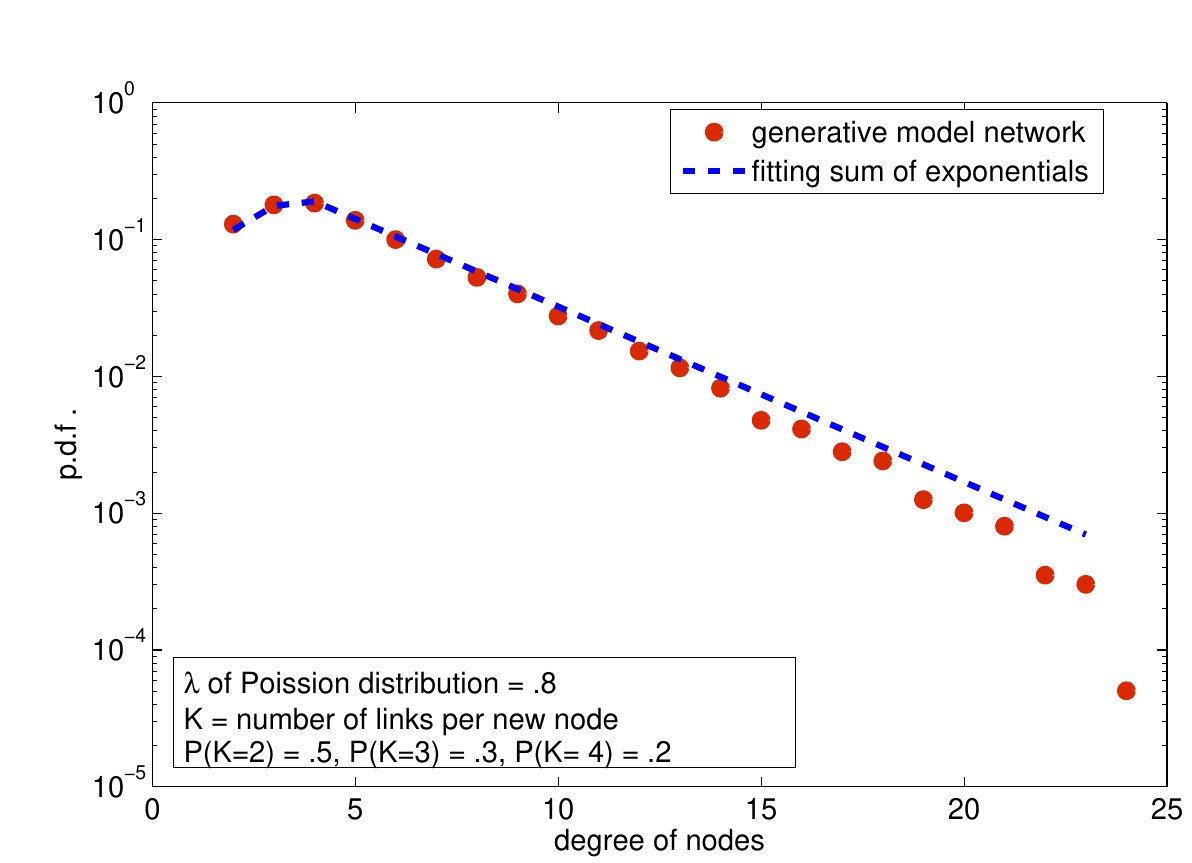}\label{fig:fit2}}%\hspace{.05cm}
\subfigure[]{\includegraphics[width=.47\textwidth,height=.45\textwidth]{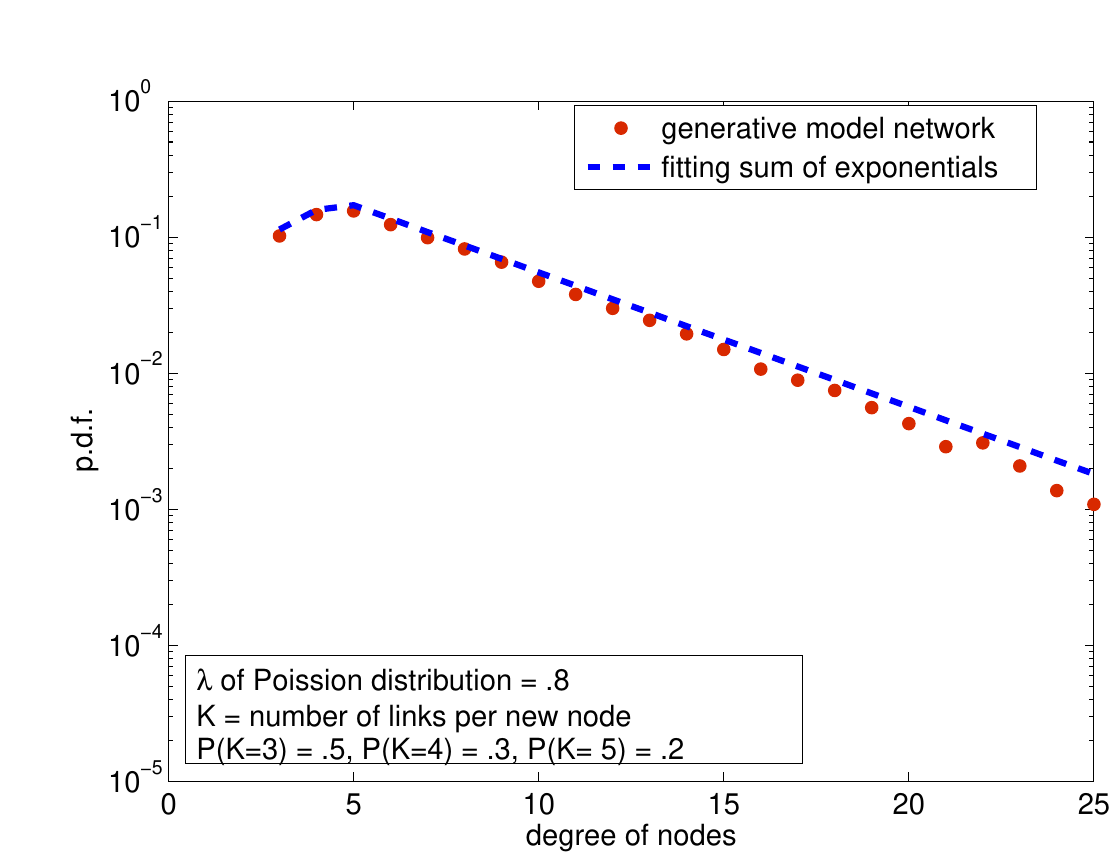}\label{fig:fit3}}%\hspace{.05cm}
\vspace{-.25cm}
\caption{Fitting sum of shifted exponential p.d.f.s from Equation (\ref{varialbek}) to the node degree distribution produced by generative model with parameter $K$ being an integer valued variable with known distribution}
%\label{fig:degreedist}
\vspace{-3mm}
\end{figure}

Thus, a mean-field approximation converges to the actual degree distribution. To elucidate our analysis, we present Figs. \ref{fig:fit3} and \ref{fig:fit3} which include the degree distribution for two networks given by our generative model with parameter $K$ taking 3 possible values with known probabilities. We see that the degree distribution in both cases is correctly fit by the expression given in Equation (\ref{varialbek}). This demonstrates the accuracy of the mean-field result for our generative model. Further it is worth noting that the nodal degree distribution of the generative model has an initial rising component followed by an exponential tail, similar to the common trend observed in degree distributions of real power grid networks. In the next section, we use our generative model to fit real power grid data.
%\begin{figure}[ht]
%\centering
%\includegraphics[width=0.49\textwidth]{fig2a.pdf}
%\caption{Fitting exponential p.d.f. to the degree distribution for fixed $K$ in radius 20 }
%\label{fig2}
%\end{figure}

\section{Comparing Real Power Grids with Generative Model}
\label{sec:real fit}
\begin{figure}[!bt]
\centering
\subfigure[]{\includegraphics[width=.47\textwidth,height=.45\textwidth]{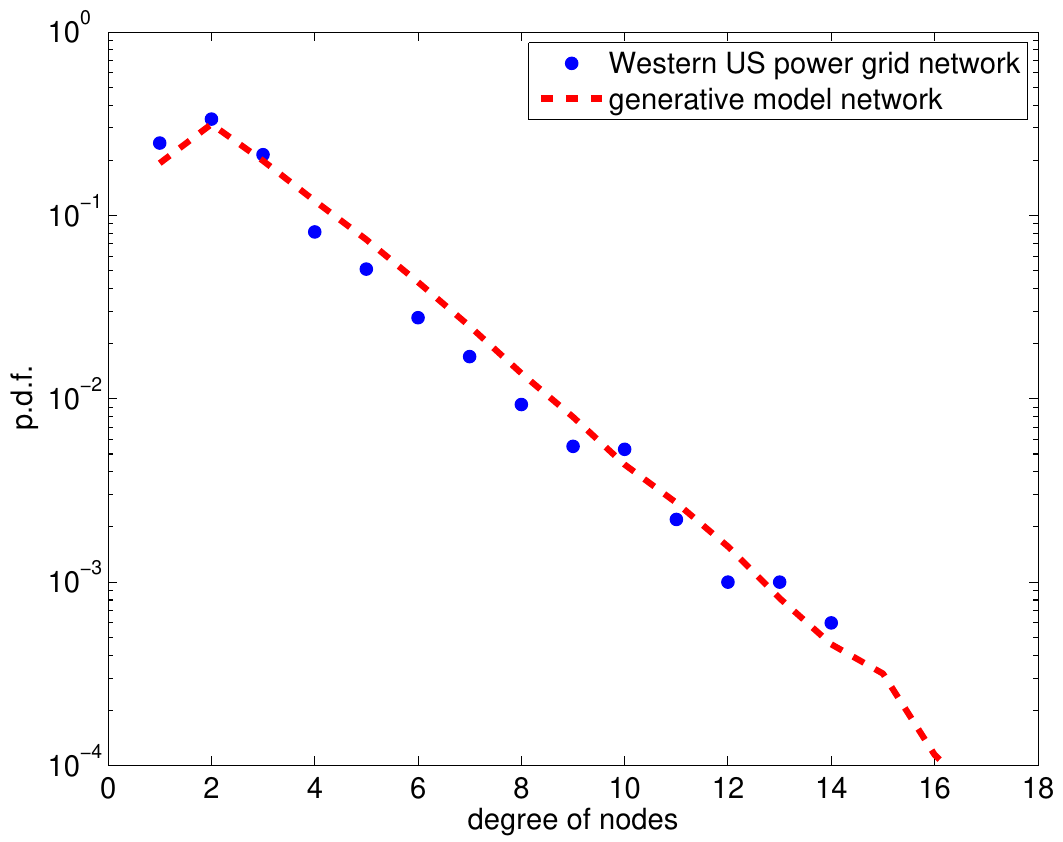}\label{fig:fitexactwestern}}%\hspace{.05cm}
\subfigure[]{\includegraphics[width=.47\textwidth,height=.45\textwidth]{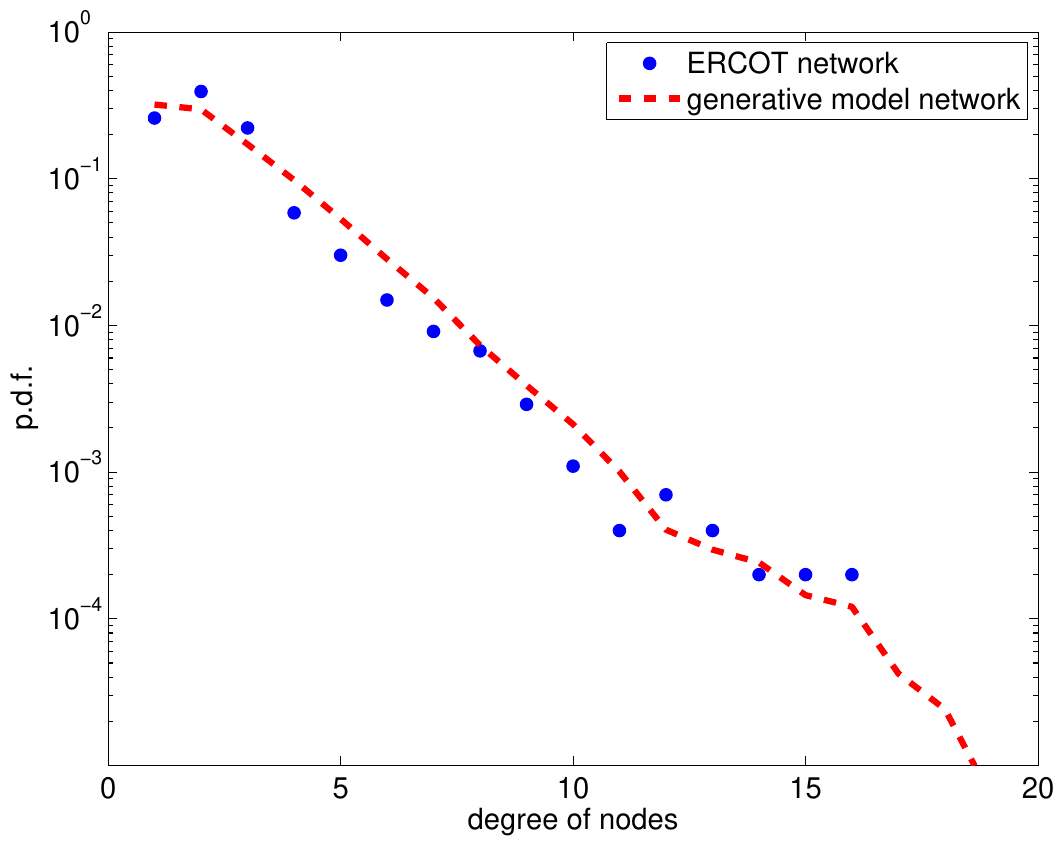}\label{fig:fitexacttexas}}%\hspace{.05cm}
\vspace{-.25cm}
\caption{Node degree distribution in real power grid and corresponding similar sized network given by generative model for (a) Western US power grid (b) ERCOT power grid.}
%\label{fig:degreedist}
\vspace{-3mm}
\end{figure}

We look at the two power grid networks described in Section \ref{sec:emp} and model them using our generative framework. To model the Western US electric power grid ($4941$ nodes and $6594$ edges), the parameter $K$ in the generative model is given values of $1$, $2$ and $12$ with probabilities $.54$, $.45995$ and $.00005$ respectively. Fig.~\ref{fig:fitexactwestern} shows the comparison between the degree distribution of the Western US electric power grid and the average degree distribution of a network with equal number of nodes simulated using our generative model. Similarly, we model the ERCOT power grid ($5514$ nodes and $6522$ edges) using our generative model where parameter $K$ takes values of $1$, $2$ and $12$ with probabilities $.75$, $.24994$ and $.00006$ respectively. Fig.~\ref{fig:fitexacttexas} shows the comparison of degree distributions for this case . Note that in either case, the support of $K$ has a size of $3$, where $K$ takes the largest value ($>10$) with very lower probability ($<10^{-3}$). This large value in $K$'s support models the few critical nodes in the grid that have degrees much greater than the remaining nodes.

The preceding theoretical analysis, simulation results and fitting with real grid data signify that our intuitive generative framework is able to model the distribution of nodal degrees while preserving mathematical tractability. Next we look at other important parameters that affect the grid's functioning and security and compare their trends seen in real data with our generative model.

\section{Diameter of the Network}
\label{sec:diameter of the proposed model}
\begin{figure}[!bt]
\centering
\subfigure[]{\includegraphics[width=.45\textwidth,height=.45\textwidth]{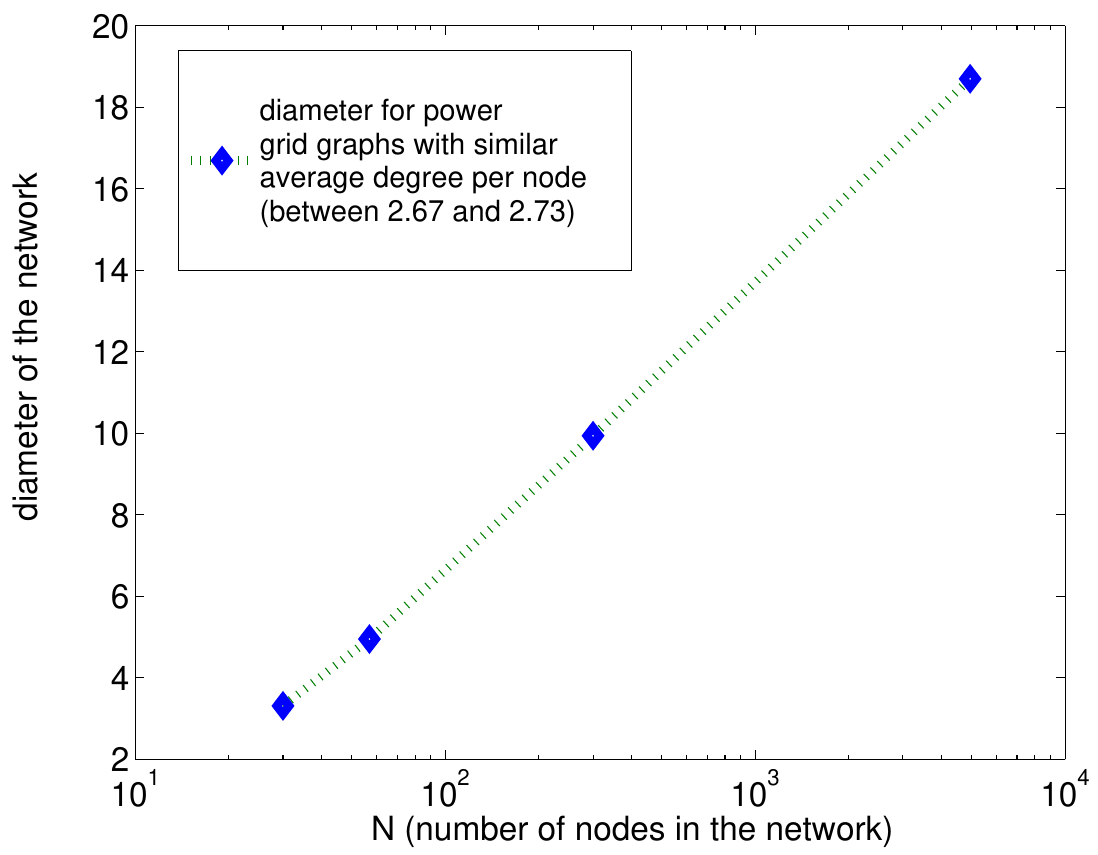}\label{fig:finaldiameter}}%\hspace{.05cm}
\subfigure[]{\includegraphics[width=.45\textwidth,height=.45\textwidth]{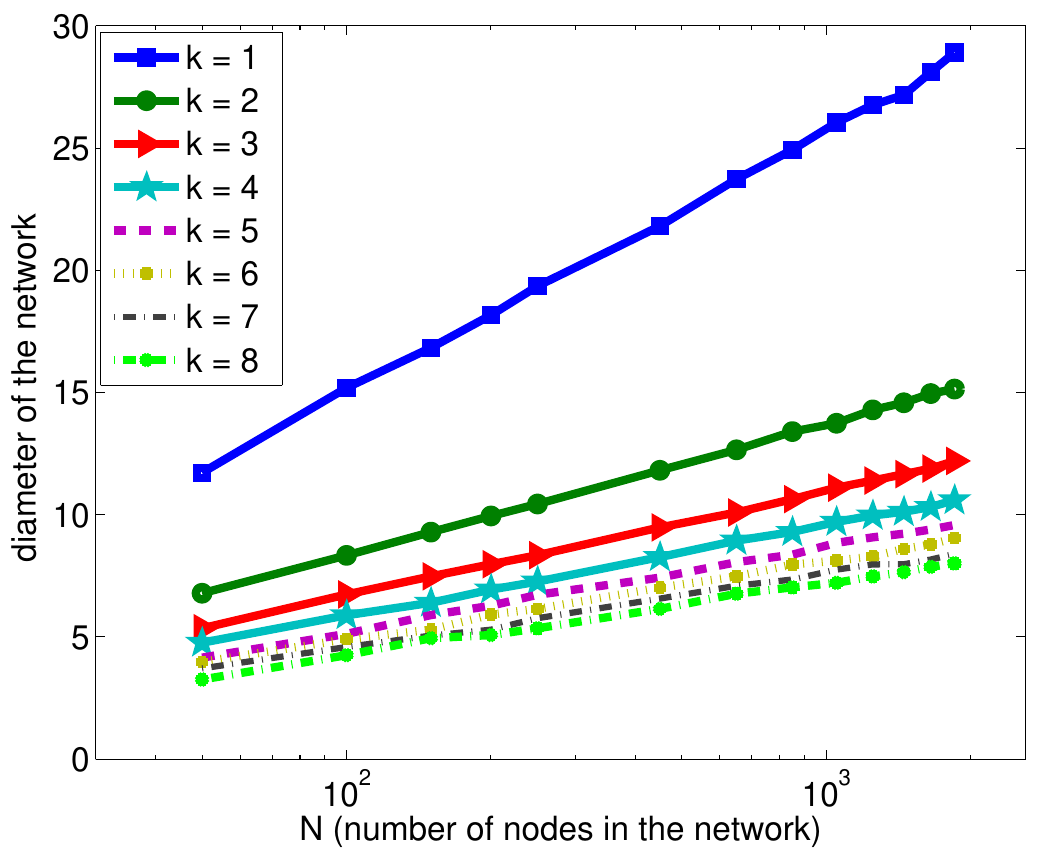}\label{fig:maxdiameter}}%\hspace{.05cm}
\vspace{-.25cm}
\caption{Diameter of the network v/s size of the power grid for (a) real power grids  (b) generative model networks.}
\vspace{-3mm}
\end{figure}

The diameter of a graph is the length of the longest route between any two nodes in the graph. In power grids, the diameter affects the probability of fragmentation following a directed or random adversarial attack on the network nodes and edges \cite{albert2004exp2,betweenness1}. This is observed as diameter of the network is related to the prevalence of nodes with high degree critical for connectivity in the grid. In general, the scaling of the diameter of the power grid network with the number of nodes is hard to quantify due to varying topological constraints and geographical boundaries that affect connectivity in the network. Fig. \ref{fig:finaldiameter} presents the average diameter observed in power grids with comparable average nodal degree as presented in \cite{hines2011,wang2010}. Note that the average diameter of the network scales with the logarithm of the number of nodes. The diameter of networks given by our generative model for different values of $N$ (number of nodes) and $K$ (number of connections per incoming node) is shown in Fig.~\ref{fig:maxdiameter}. We observe a logarithmic scaling between the diameter and the network size for our generative model, that is similar to observations in power grids.

\section{Betweenness Centrality of the Generative Model}
\label{sec:Betweenness Centrality of the proposed model}
In this section, we compare the betweeness centrality of graphs given by our generative model with that of available power grid graphs. Betweenness centrality for a node $i$ (denoted by $l_i$) or an edge $e_{im}$ (denoted by $l_{im}$) in a graph measures the number of shortest paths (with minimum number of hops) in the graph that pass through that node or edge respectively \cite{crucittia2004}. For an undirected graph with $N$ nodes, betweenness centralities of every node and edge are given by:
\begin{align}
&l_i = \sum_{j,k \neq i}\frac{n_{jk}(i)}{n_{jk}} \text{~~~betweenness centrality for node~} i \\
&l_{im} = \sum_{j,k \neq i,m}\frac{n_{jk}(im)}{n_{jk}} \text{~~~betweenness centrality for edge~} e_{im}
\end{align}
Here $n_{jk}$  is the total number of shortest paths between nodes $j$ and $k$ while $n_{jk}(i)$ is the number of shortest paths between $j$ and $k$  that include node $i$. Similarly, $n_{jk}(im)$ denotes the number of shortest paths between $j$ and $k$ that pass through edge $e_{im}$.

Higher value of betweenness centrality for a node or an edge indicates its greater relevance to the robustness of the network. In particular, studies show that an attack on nodes with the largest betweenness centrality values creates worse blackouts and loss of connectivity in a power grid than an attack on nodes with the highest degrees \cite{crucittia2004, betweenness}. Further, \cite{homes2002} mentions the positive correlation between the betweenness centrality of an edge in the grid with the amount of load flowing on the edge. This enables the use of betweenness centrality in the analysis of cascading outages in the power grid \cite{westernusacascade}.

\begin{figure}[!bt]
\centering
\subfigure[]{\includegraphics[width=.45\textwidth,height=.40\textwidth]{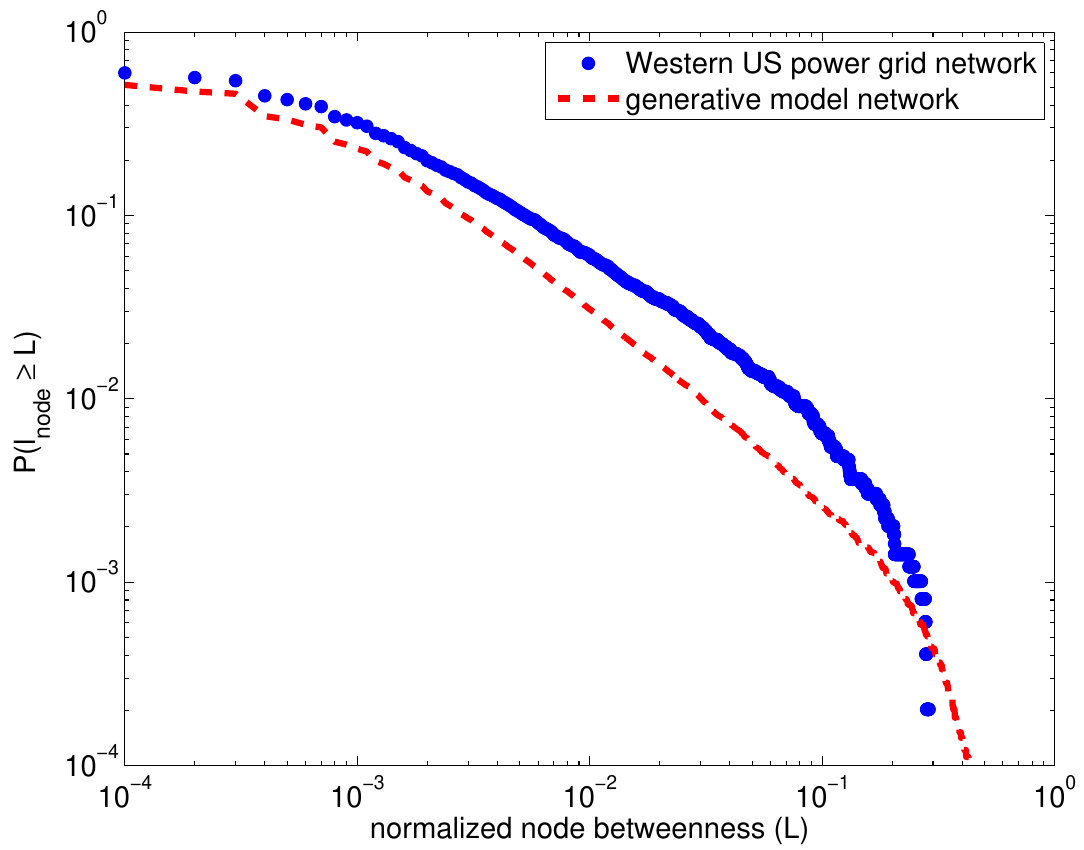}\label{fig:westernusabetweennessnode}}%\hspace{.05cm}
\subfigure[]{\includegraphics[width=.45\textwidth,height=.40\textwidth]{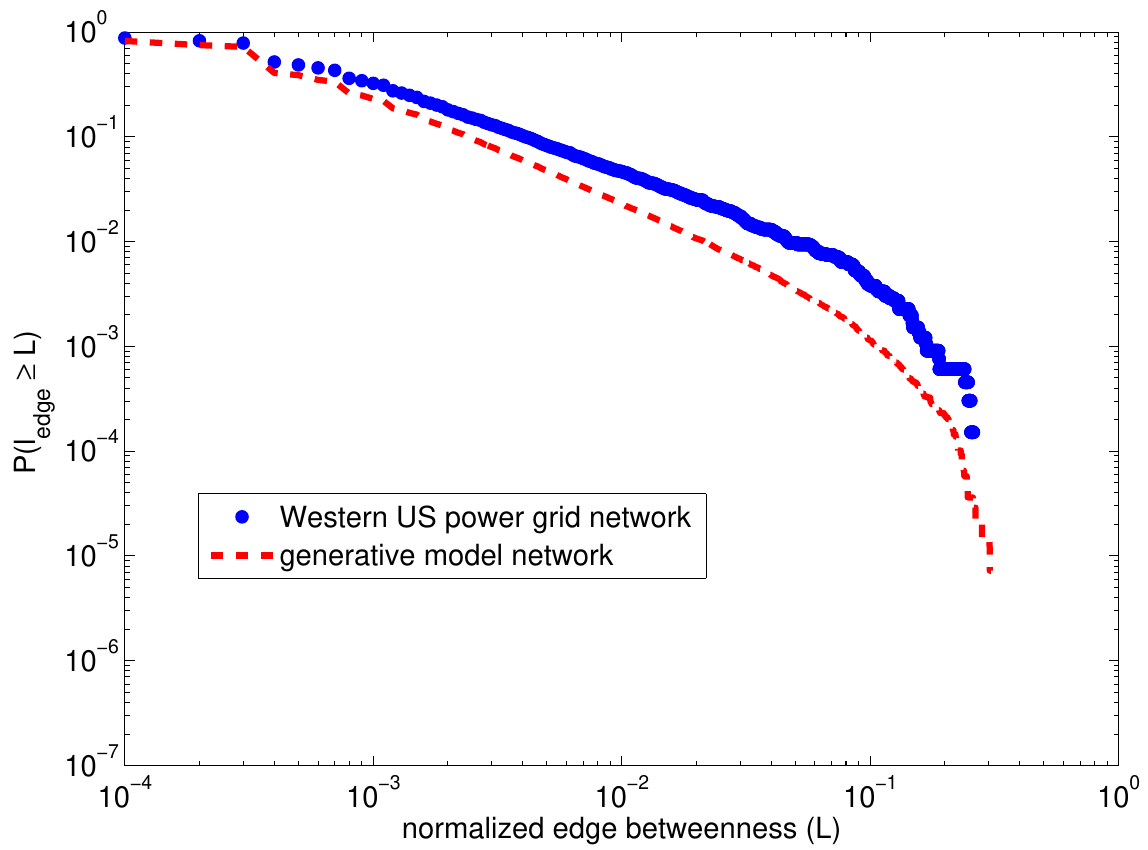}\label{fig:westernusabetweennessedge}}%\hspace{.05cm}
\vspace{-.25cm}
\caption{Comparison between Western US power grid and similar sized network given by generative model for (a) Node betweenness centrality  (b) Edge betweenness centrality.}
\end{figure}

\begin{figure}[!bt]
\centering
\subfigure[]{\includegraphics[width=.45\textwidth,height=.40\textwidth]{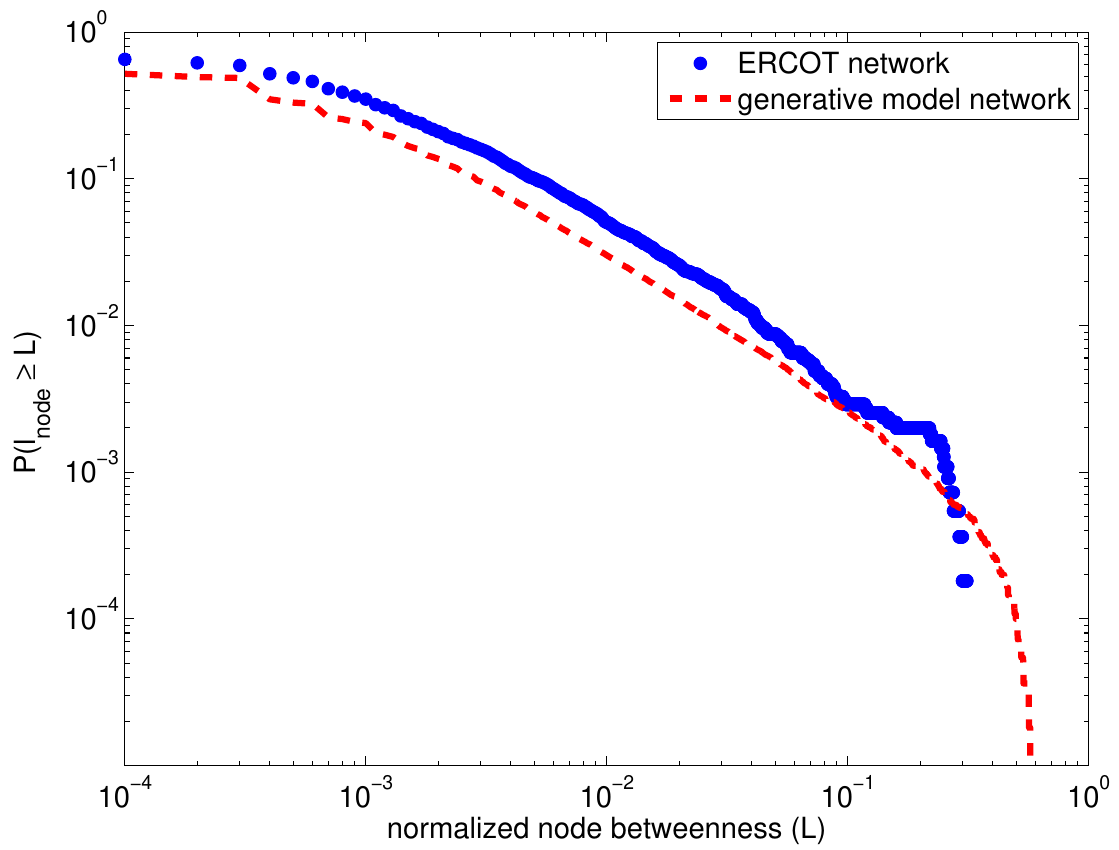}\label{fig:texasgridbetweennessnode}}%\hspace{.05cm}
\subfigure[]{\includegraphics[width=.45\textwidth,height=.40\textwidth]{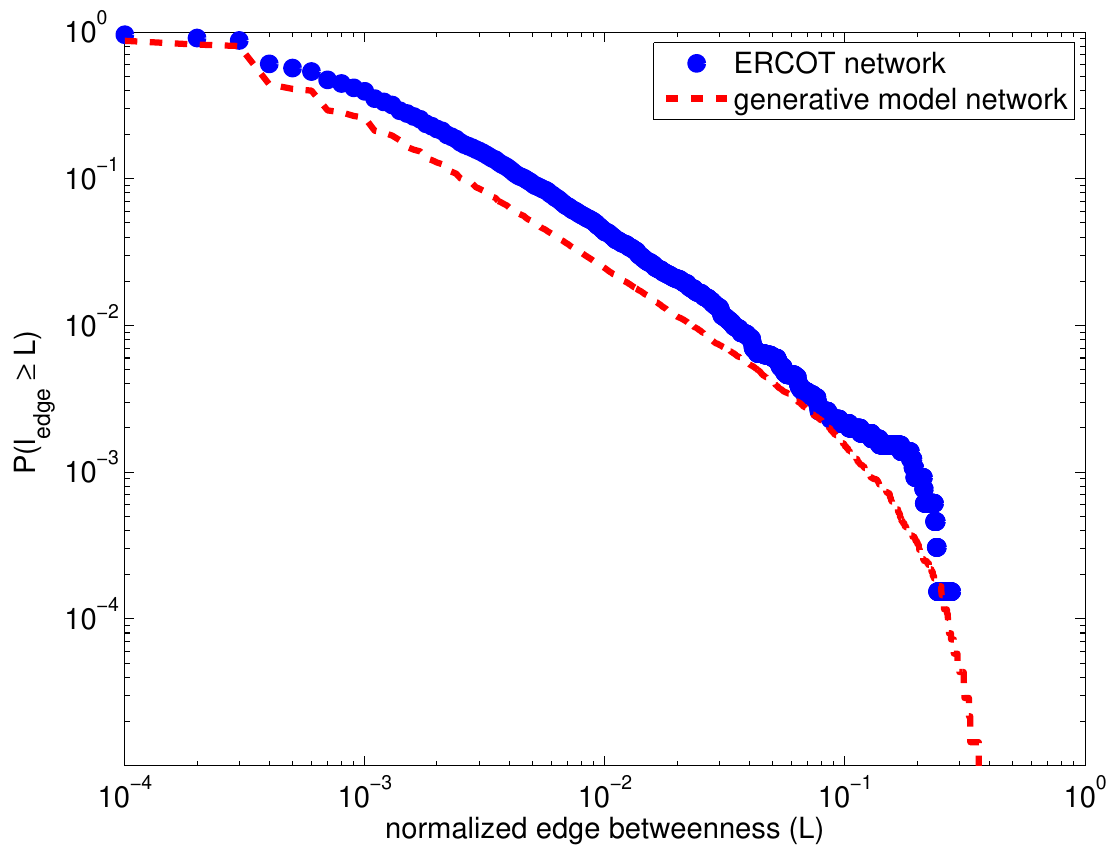}\label{fig:texasgridbetweennessedge}}%\hspace{.05cm}
\vspace{-.25cm}
\caption{Comparison between ERCOT power grid and similar sized network given by generative model for (a) Node betweenness centrality  (b) Edge betweenness centrality.}
\end{figure}

We plot the p.d.f.s of node and edge betweenness centralities ($L$) for Western US power grid in Figs.~\ref{fig:westernusabetweennessnode} and \ref{fig:westernusabetweennessedge}. For each betweenness centrality measure, we also plot the average p.d.f. obtained from a similar sized network given by our generative model. It is evident from the plots that our generative model accurately predicts both the node and edge betweenness centralities for the Western US power grid. Next we consider the ERCOT power grid. We show the p.d.f.s of its edge and node betweenness centralities in Figs.~\ref{fig:texasgridbetweennessnode} and \ref{fig:texasgridbetweennessedge} respectively and compare them with the corresponding centrality measures of networks given by our generative model. We observe that our generative model induces centrality characteristics that closely resemble that of the ERCOT grid. Note that, to model the two power grids, our generative model uses values of parameter $K$ mentioned in Section \ref{sec:real fit}.

\section{Node Clustering of the Generative Model}
\label{sec:Node Clustering of the proposed model}
\begin{figure}[!bt]
\centering
\subfigure[]{\includegraphics[width=.45\textwidth,height=.40\textwidth]{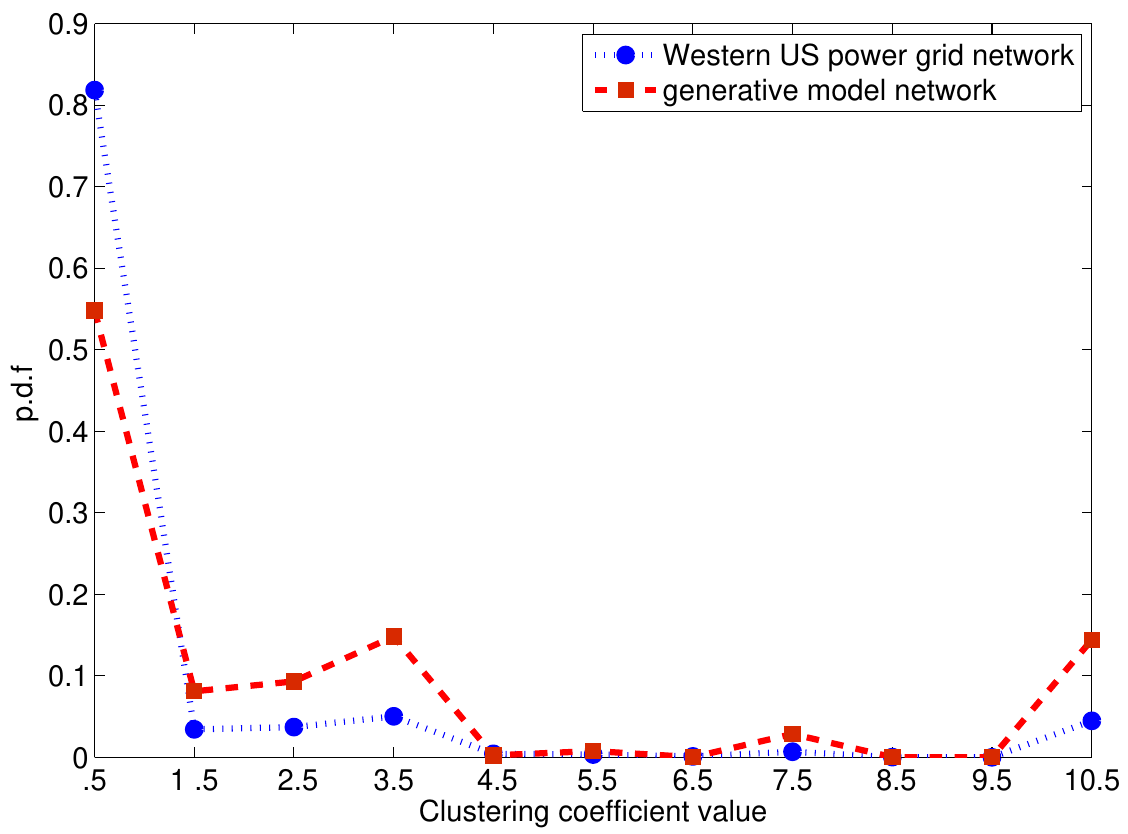}\label{fig:clusterfitwestern}}\hspace{.05cm}
\subfigure[]{\includegraphics[width=.45\textwidth,height=.40\textwidth]{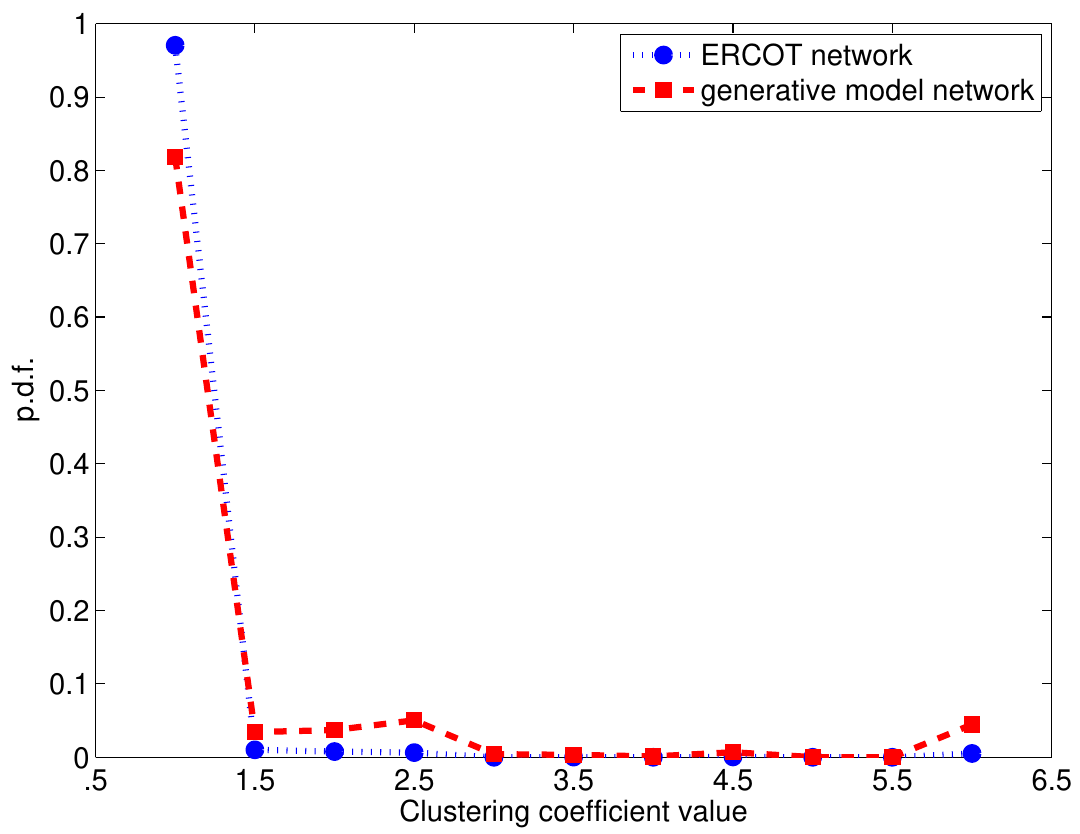}\label{fig:clusterfittexas}}%\hspace{.05cm}
\vspace{-.25cm}
\caption{Comparison of clustering coefficient between real power grids and generative model networks. (a) Western US power grid (b) ERCOT power grid}
\end{figure}

Now, we look at the clustering of nodes in networks simulated with our generative model. This refers to the phenomenon where neighbors of a node have links with one another leading to the formation of clusters. One of the distinguishing features of social and physical networks is that the clustering observed in them is higher than that of Erd$\ddot{o}$s-R$\acute{e}$nyi random graphs. Here, we focus on the node clustering coefficient and compare its distribution for power grid graphs and those generated by our generative model. Mathematically, the clustering coefficient for a node in graph $G$ is defined as follows:

\begin{eqnarray}
C_i = \pi_i /\Pi_i
\end{eqnarray}
where $\pi_i$ is the number of edges that exist between neighbors of node $i$ and $\Pi_i$ are the total number of edges that can be formed using $i$'s neighbors. Clearly, $\Pi_i$ is given by $d_i(d_i -1)/2$ where $d_i$ is the degree of node $i$. The maximum value of the clustering coefficient is $1$, and attained for nodes that form cliques (complete clusters) with their neighbors. Fig.~\ref{fig:clusterfitwestern} shows the distribution of clustering coefficients in the Western US power grid and that obtained on average for similar sized networks produced by our generative model. The comparison for the clustering coefficient distribution of the Texas power grid under ERCOT is given in Fig.~ \ref{fig:clusterfittexas}. Note that the clustering coefficient given by our generative model is similar to that observed in real grids, though the performance is more pronounced for the ERCOT power grid.

\section{Eigen-spread of the Generative Model}
\label{sec:eigen spread in the proposed model}
Eigenvalues of the adjacency matrix of a graph is a key connectivity parameter that affects the speed at which network processes like infection (discussed in later sections) and topological vulnerabilities spread from one part of the network to another. Here, we plot the magnitude of the largest five largest eigenvalues of the adjacency matrix of power grid networks and compare them with those observed on average in similar sized networks given by our generative model. Though the general trend is similar, we note that our model is not able to simulate the largest eigenvalue in the both the considered power grids. We plan to study the structural changes needed to improvise the modeling the largest eigenvalue in the power grids in future work.

\begin{figure}[!bt]
\centering
\subfigure[]{\includegraphics[width=.45\textwidth,height=.40\textwidth]{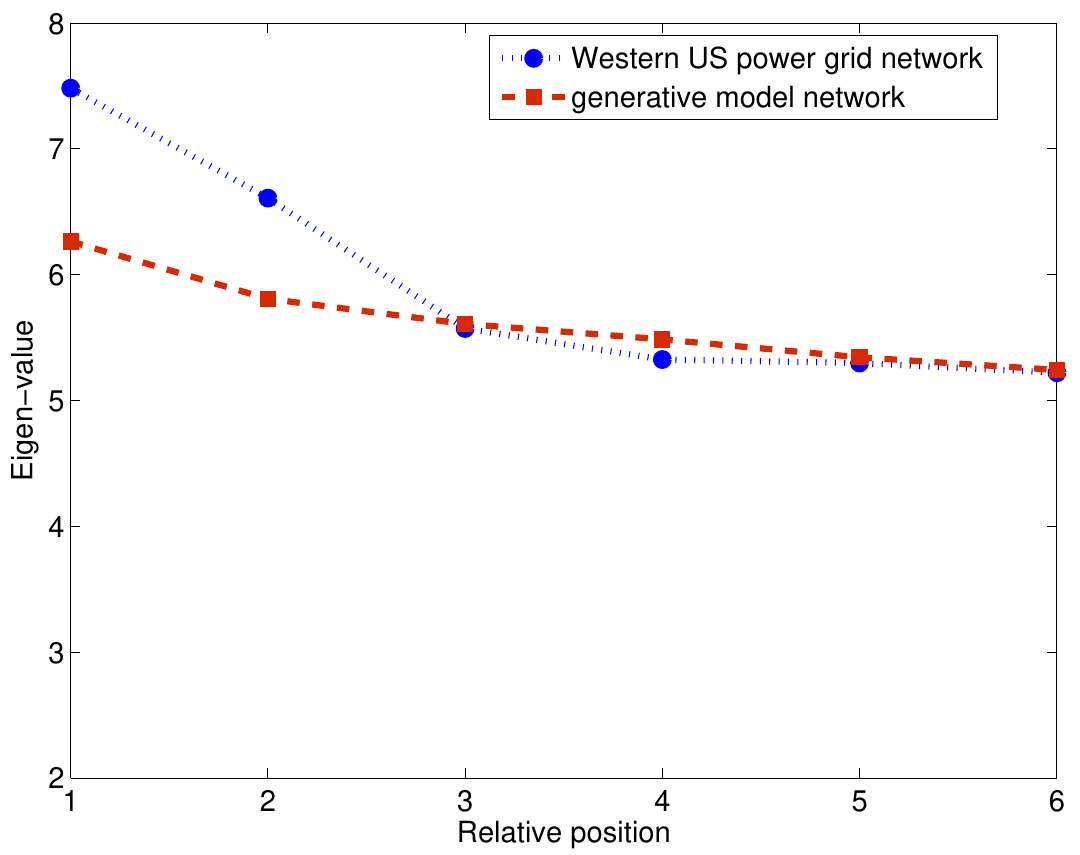}\label{fig:eigenfitwestern}}\hspace{.05cm}
\subfigure[]{\includegraphics[width=.45\textwidth,height=.40\textwidth]{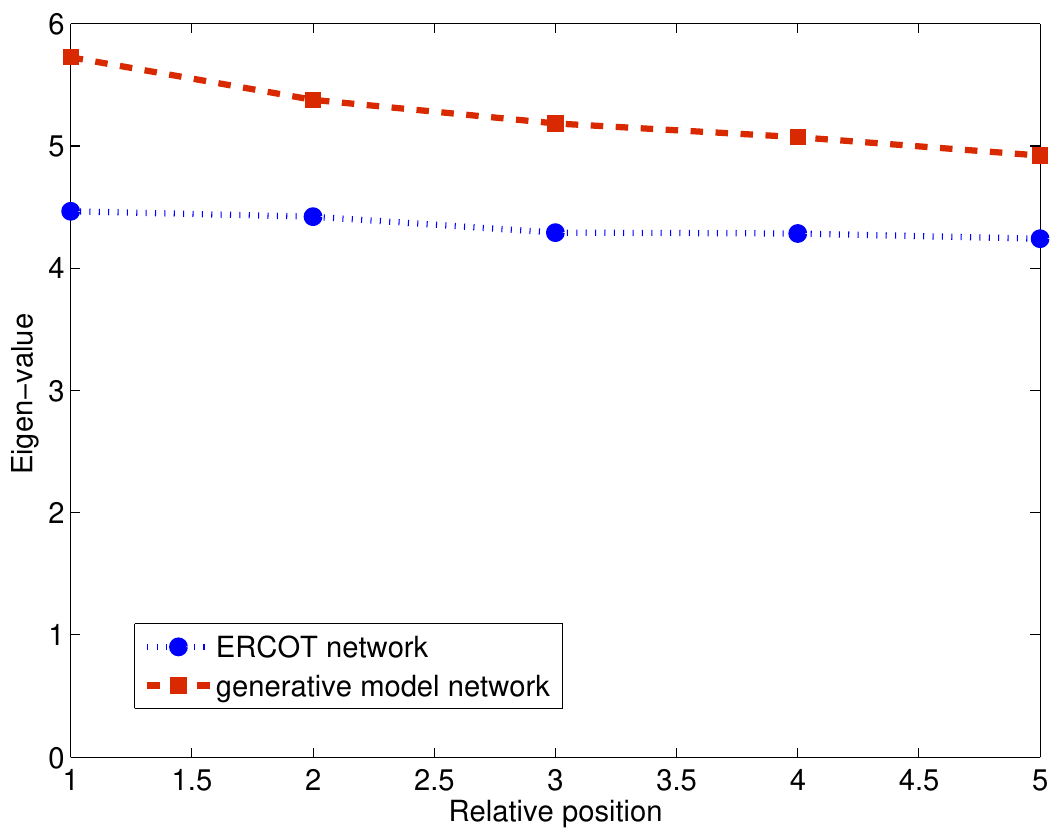}\label{fig:eigenfittexas}}%\hspace{.05cm}
\vspace{-.25cm}
\caption{Comparison of spread of eigenvalues between real power grids and generative model networks. (a) Western US power grid (b) ERCOT power grid}
\end{figure}

\section{Vulnerability of the Generative Model}
\label{sec:Vulnerability of the proposed model}
Structural features of the power grid, in particular the parameters discussed in the preceding sections, help determine the topological vulnerability of the grid to random and intentional attacks by disruptive agents. Such attacks can, in principle, disrupt the secure operation of the grid and delivery of electricity to end users. In extreme cases, even a small disturbance inserted in the system from outside can propagate to the rest of the network and cause a cascading outage. In the past, topological failures in independent power grids as well as interdependent infrastructure networks have been studied using different failure propagation models \cite{interdependent}. Our interest lies in understanding if our generative model is capable of simulating failure propagation in real grids.

We consider two models of failure propagation in the power grid here: the SIS (Susceptible-Infected-Susceptible) model and the SIR(Susceptible-Infected-Removed) model  \cite{infection1, epidemiccorr}. These models were originally derived for studying infection propagation on biological networks. In the power grid, the same models have been used to study the propagation of frequency oscillations and disturbances \cite{frequencyoscillations, deka_isgt2015}. The SIS mode \cite{infection1} comprises of nodes in two states : susceptible (S) or infected (I). While an infected node returns to susceptible state with probability $\delta$, a susceptible node gets the infection from an infected neighbor with probability $\beta$. Here, $\delta$ and $\beta$ are parameters of the infection. Similarly, the SIR model \cite{infection1} has nodes in three states: susceptible (S), infected (I) and removed (R), differing from the SIS model in the fact that a infected node enters the removed state with probability $\gamma$ and stays in that state.

\begin{figure}[!bt]
\centering
\subfigure[]{\includegraphics[width=.45\textwidth,height=.40\textwidth]{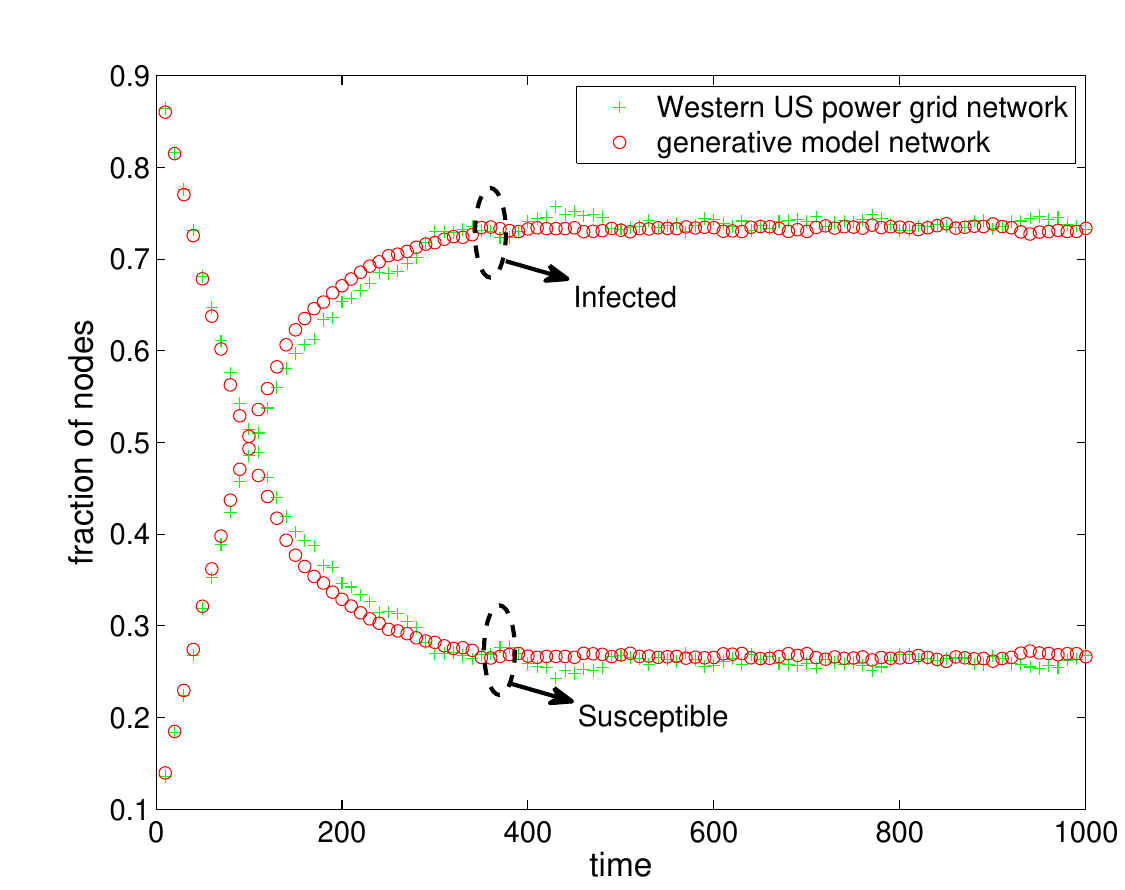}\label{fig:SISwestern}}\hspace{.05cm}
\subfigure[]{\includegraphics[width=.45\textwidth,height=.40\textwidth]{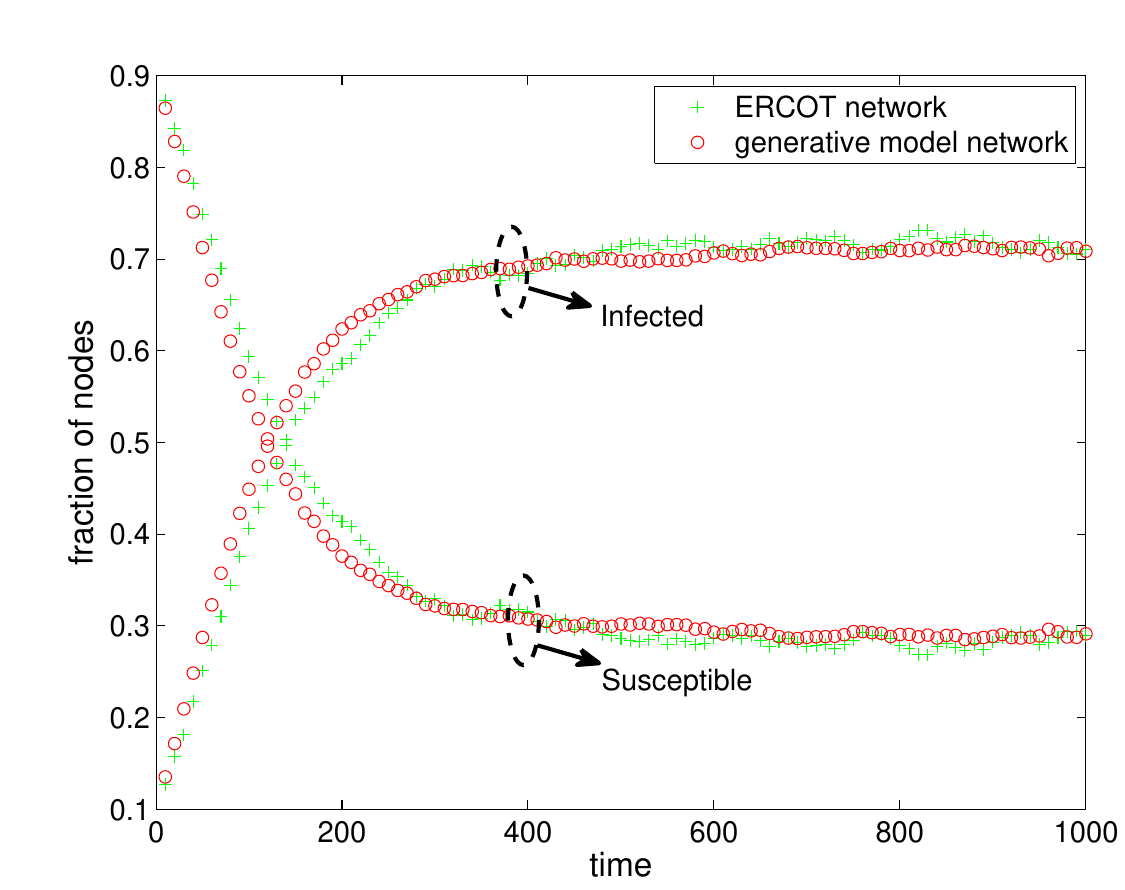}\label{fig:SIStexas}}%\hspace{.05cm}
\vspace{-.25cm}
\caption{Comparison of SIS infection propagation in real power grids and generative model networks. (a) Western US power grid (b) ERCOT power grid}
\end{figure}

\begin{figure}[!bt]
\centering
\subfigure[]{\includegraphics[width=.45\textwidth,height=.40\textwidth]{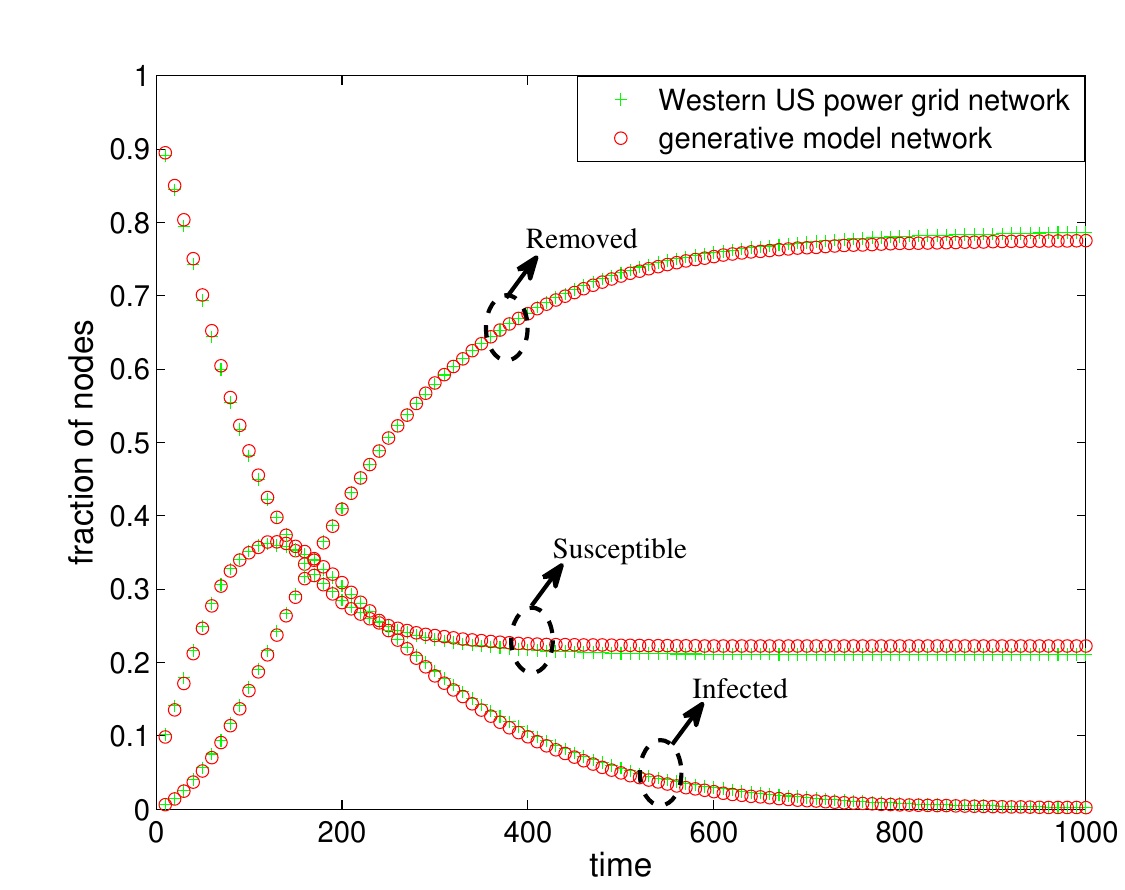}\label{fig:SIRwestern}}\hspace{.05cm}
\subfigure[]{\includegraphics[width=.45\textwidth,height=.40\textwidth]{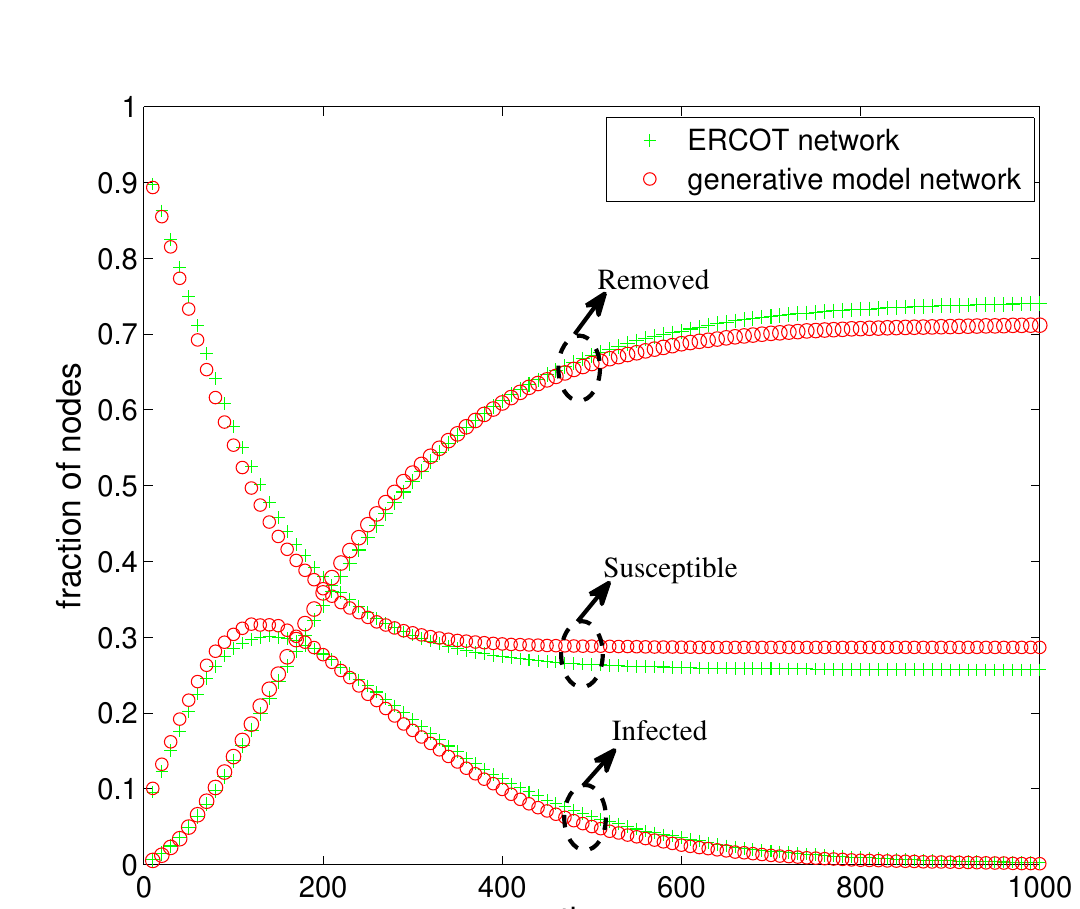}\label{fig:SIRtexas}}%\hspace{.05cm}
\vspace{-.25cm}
\caption{Comparison of SIR infection propagation in real power grids and generative model networks. (a) Western US power grid (b) ERCOT power grid}
\end{figure}

We consider a particular case of SIS and SIR infection propagation specified by the values of their respective parameters. Figs. \ref{fig:SISwestern} and \ref{fig:SIStexas} show averaged results for SIS simulation on the Western American Power Grid and the ERCOT network. In either figure, average results derived from simulations on similar-sized networks formed by our generative model are also plotted for comparison. Next, Figs.~\ref{fig:SIRwestern} and \ref{fig:SIRtexas} show the results of simulating SIR infection on the Western American Power Grid and ERCOT grid and on similar sized networks generated by our generative model. The similarity between the infection propagation curves in real and generated networks highlights the potential use of our generative model in designing grid resilience measures.


\section{Conclusion}
\label{sec:results and conclusion}
This paper studies the topological characteristics of large power grids, specifically the Western US and ERCOT power grid, and presents an intuitive and analytical generative model for their physical networks. The model simulates the growth of power grids as a temporally evolving point process where cost effective transmission lines are built to connect new nodes for reliable delivery of electricity. Through analysis and simulations, it is shown that the nodal degree distribution produced by the generative model is a weighted sum of shifted exponentials that matches with the observed degree distributions in several American and European power grids. The efficacy of the generative model is further demonstrated by comparing several graph features, namely graph diameter, node and edge betweenness centralities, spread of eigen-values and clustering coefficients, of real grids with that of similar sized simulated networks. The similarities in characteristics signify that the generative model can be used to theoretically study common topological attributes in large power grids and derive analytical results to quantify and improve their functioning and robustness. As an elucidatory example, dynamics of SIS and SIR infection propagation are conducted on real grid topologies and similar sized artificial networks produced by the generative model. Simulations verify that the generative model provides a good match for the temporal dynamics of infection propagation in the real grids. Incorporating additional features of the power grid into our generative model without sacrificing mathematical tractability of the model is challenging. This is the focus of our future research in this domain.

\section*{Acknowledgement}
The authors thank Prof. Ross Baldick of The University of Texas at Austin for insightful discussions and for providing the ERCOT power grid data set.

\end{document}